\documentclass[twocolumn]{aastex631}

\usepackage{amssymb,amsmath}
\hypersetup{colorlinks=true,linkcolor=blue,citecolor=blue,urlcolor=magenta}

\begin{document}

\title{Relativistic Collisionless Shocks in Inhomogeneous Magnetized Plasmas}
\author[0000-0003-3433-0772]{Camilia Demidem}
\email{camilia.demidem@colorado.edu}
\affiliation{Nordita, KTH Royal Institute of Technology and Stockholm University, Hannes Alfv\'{e}ns v\"{a}g 12, SE-106 91 Stockholm, Sweden}
\affiliation{JILA, University of Colorado and National Institute of Standards and Technology, 440 UCB, Boulder, CO 80309-0440, USA}

\author[0000-0002-3226-4575]{Joonas N\"{a}ttil\"{a}}
\affiliation{Center for Computational Astrophysics, Flatiron Institute, 162 Fifth Avenue, New York, NY 10010, USA}
\affiliation{Department of Physics and Columbia Astrophysics Laboratory, Columbia University, New York, NY 10027, USA}

\author[0000-0002-5767-7253]{Alexandra Veledina}
\affiliation{Department of Physics and Astronomy, FI-20014 University of Turku, Finland}
\affiliation{Nordita, KTH Royal Institute of Technology and Stockholm University, Hannes Alfv\'{e}ns v\"{a}g 12, SE-106 91 Stockholm, Sweden}

\begin{abstract}

Relativistic collisionless shocks are associated with efficient particle acceleration when propagating into weakly magnetized homogeneous media; as the magnetization increases, particle acceleration becomes suppressed.
We demonstrate that this changes when the upstream carries kinetic-scale inhomogeneities, as is often the case in astrophysical environments.
We use fully kinetic simulations to study relativistic perpendicular shocks in magnetized pair plasmas interacting with upstream density perturbations.
For amplitudes of $\delta \rho/\rho \gtrsim 0.5$, the upstream fluctuations are found to corrugate the shock front and generate large-scale turbulent shear motions in the downstream, which in turn are capable of accelerating particles.
This can revive relativistic magnetized shocks as viable energization sites in astrophysical systems, such as jets and accretion disks.
The generation of large-scale magnetic structures also has important implications for polarization signals from blazars. \smallskip

\noindent \emph{Unified Astronomy Thesaurus concepts:} Shocks (2086); Relativistic jets (1390); Computational methods (1965); High energy astrophysics (739); Plasma astrophysics (1261)
\end{abstract}


\section{Introduction} \label{sec:intro}

Relativistic collisionless shocks---shocks formed in flows propagating with bulk velocities close to the speed of light and mediated by small-scale kinetic plasma processes---are frequently observed in various astrophysical systems \citep{Bykov2012, Sironi2015review}.
Based on theoretical and numerical studies in the test particle limit, these shocks have long been suspected to be efficient particle accelerators \citep{Axford1977,Krymskii1977,Bell1978-1,Bell1978-2,BlandfordOstriker1978,Drury1983,BlandfordEichler1987,Bednarz_Ostrowski_98, 01Achterberg, Lemoine_Pelletier_03, 05Keshet}. 
This was confirmed by first-principles particle-in-cell (PIC) simulations for shocks with low magnetization parameters $\sigma$, the ratio of magnetic and particle energy densities of the upstream plasma.

At higher magnetizations, particle acceleration appears to be limited to the special case of quasi-parallel shocks, when the angle between the background magnetic field and the shock normal (in the upstream rest frame) is smaller than a critical value, inversely proportional to the shock Lorentz factor \citep{09Sironi}.
For the case of quasi-perpendicular shocks in pair plasmas, particle acceleration was found to be inhibited for magnetizations, $\sigma \gtrsim 10^{-3}$ \citep{Langdon1988,Gallant1992,09Sironi,13Sironi}.

The absence of particle acceleration from relativistic magnetized shocks in previous kinetic simulations has had important astrophysical implications.
Magnetized relativistic outflows from compact objects, such as jets in gamma-ray bursts, X-ray binaries, or active galactic nuclei (AGNs), display synchrotron and inverse Compton emission, which can be attributed to nonthermal particles, and stand as potential sources of ultrahigh-energy cosmic rays \citep{Bykov2012,Alves_Batista2019}. 
The physical processes and conditions of particle acceleration in these environments remain a subject of debate. 
In light of the results reported above, it is, however, generally considered that at high magnetization, relativistic shocks are poor particle accelerators and therefore play a minor role in the generation of nonthermal particles, compared to other phenomena such as magnetic reconnection \citep{SironiSpitkovsky2014,WernerUzdensky2021} and/or turbulence \citep{17Zhdankin,18Zhdankin,18Comisso,19Comisso,20Wong,NattilaBeloborodov21}.
Yet, spectra and timing properties of jets in AGN and X-ray binaries have been successfully explained in terms of the internal shock model, which attributes nonthermal radiation to multiple shocks along the jet \citep{Rees1978,Malzac2018}.

\begin{figure}[t]
    \centering
    \includegraphics[clip, trim=0.0cm -0.5cm 0.0cm -0.5cm, width=0.45\textwidth]{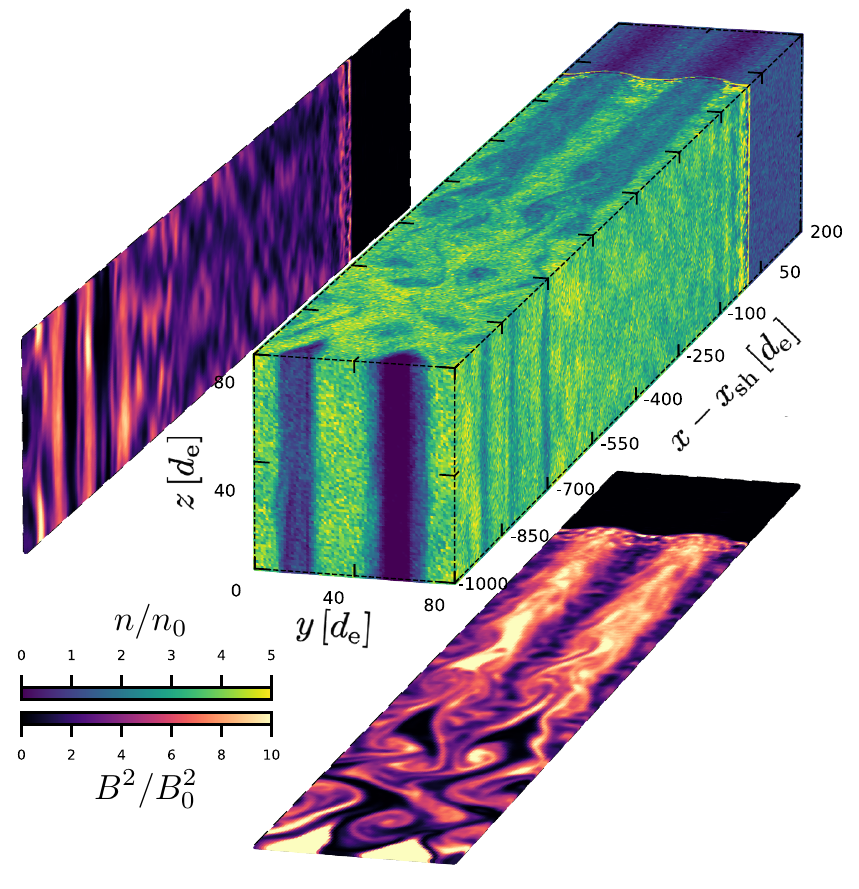}
    \caption{
    Structure of a corrugated collisionless shock located at $x\approx x_{\rm sh}$. 
    The box shows the plasma density in units of the initial mean upstream value, $n/n_0$, and the projected panels show the magnetic energy density in units of its initial upstream value, $B^2/B_0^2$. 
    }
    \label{fig:1}
\end{figure}

Importantly, the studies reporting the suppression of particle acceleration in relativistic magnetized perpendicular shocks assume that the preshock medium is initially homogeneous.
In an astrophysical or laboratory setting, however, the background medium may carry inhomogeneities in a range of scales, including kinetic ones \citep[e.g.,][]{16Chen,20Fiuza}.

Previous numerical studies of shocks interacting with perturbed media have mainly focused on the effects of the perturbations on the shock structure and downstream magnetic field amplification \citep[e.g.,][]{11Inoue,12Guo,14Mizuno,16Ji,18CD,19Tomita}. 
Some studies based on kinetic plasma simulations also reported particle acceleration for specific nonuniform magnetic field configurations, relevant for striped pulsar winds \citep{SironiSpitkovsky2011,20Cerutti}.

In this Letter, we use first-principles simulations to study relativistic shocks propagating into nonuniform-density pair plasmas with magnetization $\sigma \sim 0.1$,
associated with particle acceleration inhibition in the homogeneous case.
We show that the interaction between the shock and the upstream inhomogeneities induce electromagnetic fluctuations in the downstream, which can generate a population of nonthermal particles.

\begin{figure*}[ht]
    \centering
    \includegraphics[clip, trim=0.0cm -0.0cm 0.0cm -0.5cm, width=0.98\textwidth]{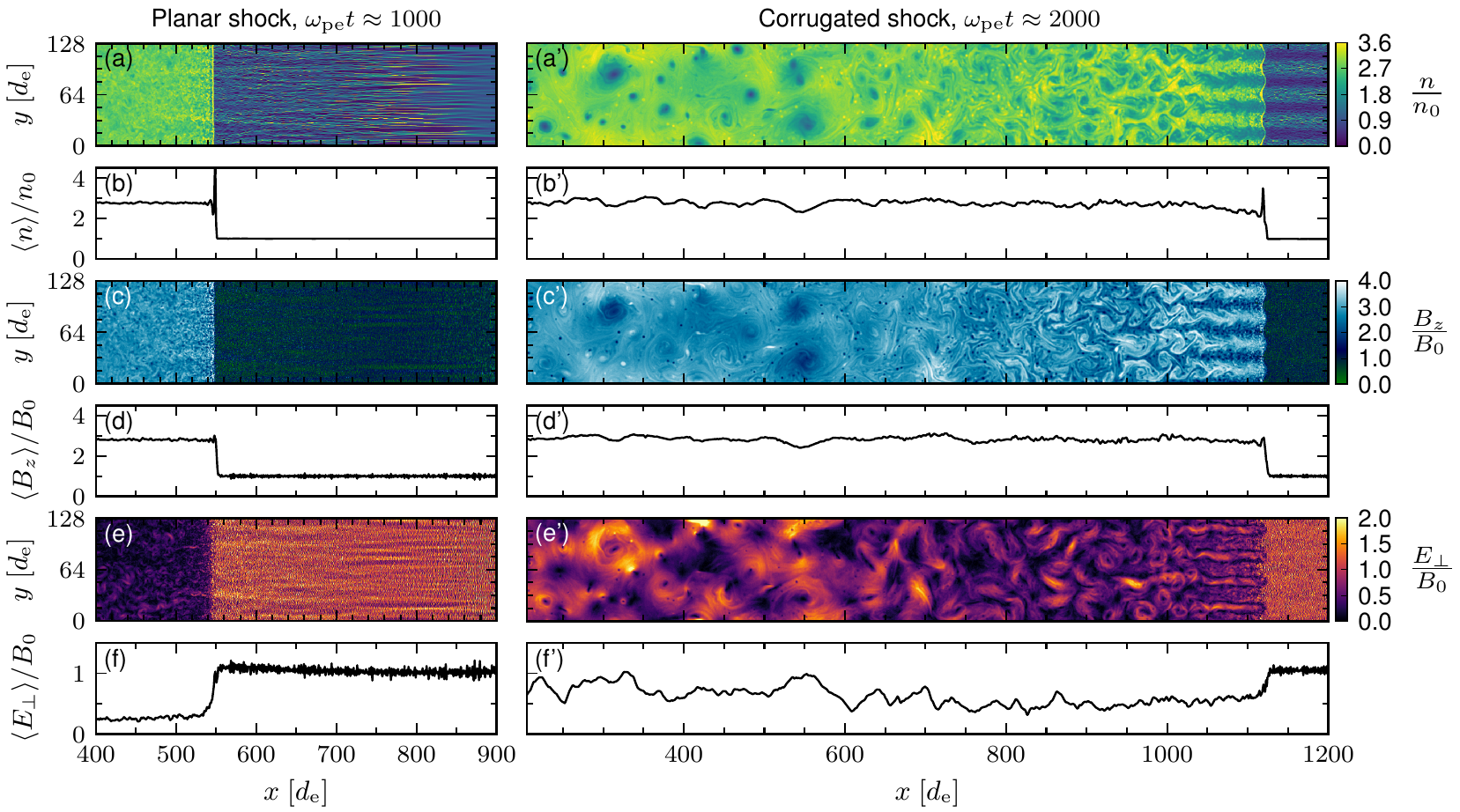}
    \caption{(a)--(f) Shock structure at $\omega_\mathrm{pe} t\approx 1000$ for a homogeneous upstream medium. (a')--(f') Shock and downstream structure at $\omega_\mathrm{pe} t\approx 2000$ for an upstream medium carrying a density wave with $\delta=0.5$, $\lambda_y=32\,d_\mathrm{e}$. The quantities displayed are, from top to bottom, the total density in units of the initial mean upstream value $n/n_0$, the out-of-plane magnetic field in units of its initial upstream value, $B_z/B_0$, and the magnitude of the in-plane electric field, $E_\perp/B_0=\sqrt{E_x^2+E_y^2}/B_0$. The 1D profiles are obtained by averaging the corresponding 2D fields over $y$.
    }
    \label{fig:shock_struct}
    \vspace{5pt}
\end{figure*}

\section{Methods} \label{sec:methods}
We use the relativistic PIC module of the {\sc Runko} framework \citep{19JN} to simulate the formation and propagation of relativistic magnetized perpendicular shocks in pair plasmas. 
Our setup for a planar shock propagating in a homogeneous upstream plasma is similar to that used in previous studies \citep{09Sironi,13Sironi,18Plotnikov}.
This ``planar'' case, validated against previous works, serves as a reference to gauge how the upstream perturbations alter the shock structure and the particle spectra.

We conducted 3D and longer 2D simulations in which the computational domain is in the $x$--$y$ plane.
Both configurations retain all three components of particle velocities and electromagnetic fields.
The simulation domain is periodic along the $y$ (and $z$ in 3D) direction and is filled with a cold plasma (of proper temperature $T_0=10^{-5}m_\mathrm{e}c^2/k_\mathrm{B}$\footnote{We have checked that the results presented below do not depend on $T_0$ for $T_0\ll m_\mathrm{e}c^2/k_\mathrm{B}$. }, where $m_\mathrm{e}$ is the electron mass, $c$ is the speed of light, and $k_\mathrm{B}$ is the Boltzmann constant) streaming along the negative $x$-direction with bulk Lorentz factor $\Gamma_0=10$. 
The left side of the domain at $x=0$ acts as an ideal conductor (reflecting the electromagnetic fields) and a hard wall (reflecting particles elastically); the right side of the domain has standard outflow boundary conditions (Figures \ref{fig:1} and \ref{fig:shock_struct}).
Following the interaction between the two counterpropagating beams of incoming and reflected particles, a shock forms and propagates in the positive $x$-direction in the wall/simulation rest frame (which approximately coincides with the downstream frame). 
All physical quantities are evaluated in this frame unless specified otherwise.

The background magnetic field in our simulations is oriented along the $z$-axis, $\boldsymbol{B}_0 = B_0 \hat{\boldsymbol{z}}$, corresponding to an out-of-plane direction for the 2D simulations.
Its magnitude is parameterized by the magnetization (invariant under a Lorentz boost along $x$), $\sigma_0=B_0^2/(4\pi n_0 \Gamma_0 m_\mathrm{e} c^2)$, where $n_0$ is the initial total particle number density of the unperturbed plasma.
In this Letter, we present results for $\sigma_0 = 0.1$ for which the downstream particle distribution has previously been reported to be thermal in the planar case \citep{Langdon1988,Gallant1992,09Sironi,13Sironi}. 
The motional electric field, $\boldsymbol{E}_0=-\boldsymbol{\beta}_0\times \boldsymbol{B}_0 = -\beta_0 B_0 \hat{\boldsymbol{y}}$ (where $\beta_0=(1-\Gamma_0^{-2})^{1/2}$), is imposed to cancel the particle drift motion in the upstream. 
We use the computationally less expensive 2D simulations to probe the long-term properties of the shocks up to at least $\omega_\mathrm{pe} t \approx 2500$; $\omega_\mathrm{pe} = \sqrt{4\pi e^2 n_{\rm e}/\Gamma_0 m_{\rm e}}$ is the electron plasma frequency ($n_{\rm e}=n_0/2$ is the electron number density).
For planar shocks, such setups capture the essential features of the shock structure and the particle dynamics of full 3D simulations \citep{13Sironi}.

The corrugated shocks are initialized using harmonic density perturbations in the upstream plasma.
The wavevector of the perturbations lies in the $x$--$y$ plane, $\boldsymbol{k} = (k_x, k_y, 0)$, and
the initial number density at the position $(x,y)$ is 
\setlength{\belowdisplayskip}{5pt} \setlength{\abovedisplayskip}{5pt} 
\begin{equation}
  n(x,y,t=0) = n_0\left[1+ \delta \cos\left(k_x x+k_y y \right) \right],
  \label{eq:npcc}
\end{equation}
where $\delta$ is the amplitude of the perturbations.
In our fiducial configurations, $\delta=0.5$ and $k_x=k_y/100$ (the results are insensitive to the exact value of $k_x$ for $k_x\ll k_y$). The transverse wavelength in units of the (upstream) plasma skin depth $d_\mathrm{e}=c/\omega_\mathrm{pe}$ is $\lambda_y=2\pi/k_y=32\, d_{\rm e}$ or $16\, d_{\rm e}$, larger than the shock thickness, which spans a few skin depths.

The transverse size of the simulation domain is $L_y = \min(256\, d_{\rm e}, 8\,\lambda_y)$ in 2D and $80\, d_{\rm e}$ in 3D. 
We also performed larger 2D simulations with $L_y=512\,d_\mathrm{e}$ and observed no significant differences in the results.
The skin depth is typically resolved by 10 cells in 2D and 3 cells in 3D runs.
The initial number of computational particles per cell per species is on average $16/(1-\delta)$ (this results in a minimum of 16 particles, corresponding to about $2000$ particles per $d_{\rm e}^2$ in 2D and $432$ particles per $d_{\rm e}^3$ in 3D).
The simulation time step satisfies $\omega_\mathrm{pe}\Delta t \approx0.05$. 
Numerical Cherenkov radiation is mitigated by applying digital filtering with a three-point binomial stencil to the electric current; we typically use 20 passes at each time step.

\section{Shock structure} \label{sec:shock_struct}
The left side of Figure \ref{fig:shock_struct} shows the structure of the flow in 2D simulations of the planar case at $\omega_\mathrm{pe} t \approx 1000$, when the shock is well formed and the downstream particle distribution is stationary.\footnote{\label{note2} Figure \ref{fig:shock_struct} displays data from simulations with half the $y$-extension of the fiducial setups ($L_y=128\, d_{\rm e}$) and field values were binned over $d_{\rm e}/2$ to limit the transverse size of the figure and the memory usage of output data while allowing a decent rendering of the fine structure of the shock.}
The shock structure observed for the homogeneous upstream flow is consistent with previous studies of magnetized relativistic pair shocks mediated by magnetic reflection \citep{Langdon1988,Gallant1992,09Sironi,13Sironi,17Iwamoto,Iwamoto2018,PlotnikovSironi2019,BabulSironi2020,21Sironi}: it displays a sharp transition between the unshocked and shocked media coinciding with a density overshoot at $x\approx 550\, d_\mathrm{e} $ in panels (a) and (b) of the figure as well as an electromagnetic precursor and density filaments.
The downstream flow is essentially laminar with negligible electric field (panels (e) and (f)); $E_\parallel$ is at the noise level throughout the simulation box and is, therefore, not displayed. 
The magnitude of the jump in density and background magnetic field measured from the $y$-averaged profiles (panels (b) and (d)) is approximately 3 and is close to the value predicted by the Rankine-Hugoniot jump relations, assuming a 2D adiabatic equation of state \cite[e.g.,][]{18CD}. 

The right side of Figure \ref{fig:shock_struct} shows the structure of the shock and downstream when the upstream medium carries a monochromatic density wave, at $\omega_\mathrm{pe} t \approx 2000$.\textsuperscript{\ref{note2}}
Four transverse wavelengths of the incoming wave are visible on panel (a') in the region $x \gtrapprox 1100\, d_\mathrm{e}$.
The $y$-dependence of the upstream density profile induces a deformation of the shock front and the generation of downstream perturbations whose development and evolution history can be traced along the $x$-axis: 
the structures farther away from the shock front are more evolved.
The amplitude, wavelength and pulsation rate of the shock ripples are set by that of the incoming wave.
The alternation of low and large densities in the upstream imprints the same pattern in the region immediately behind the shock, on the density, the in-plane bulk velocity, the out-of-plane magnetic field, and the corresponding (motional) electric field $E_\perp$, as can be seen on Figure \ref{fig:1} and panels (a')--(c') of Figure \ref{fig:shock_struct}. 
The velocity shear between the different layers excites vortices, reminiscent of the Kelvin-Helmholtz instability, whose merging and nonlinear evolution leads to the mixing of the shocked flow and to the generation of structures with sizes much larger than the skin depth.
This process is similar to what is observed when a shock evolves in a medium with an anisotropic transverse magnetic field \citep{20Cerutti}.
The motional (in-plane) electric field, induced by the bulk velocity perturbations, has significantly higher magnitudes than the planar shock case (up to a factor $\approx 4$).
We observe the same behavior in 2D and 3D simulations (slices of density and magnetic energy density are shown in Figure \ref{fig:1}, at an earlier stage, $\omega_\mathrm{pe} t=700$):
confining the density wavevector to the $x$--$y$ plane and taking a large background magnetic field along $z$ makes the shock and downstream structures invariant along the $z$-axis.

We note that the shock structure is different when considering a background magnetic field lying in the same plane as the perturbation wavevector (i.e, $\boldsymbol{B}_0$ along $y$). 
In this case, the particles can indeed quickly homogenize the plasma behind the shock front by sliding along the magnetic field lines and filling the underdense regions.
2D simulations with an in-plane magnetic field configuration show little difference with the unperturbed case at late times; we have therefore focused on configurations with $\boldsymbol{B}_0$ along $z$.

\begin{figure}[t]
    \centering
    \includegraphics[width=0.47\textwidth]{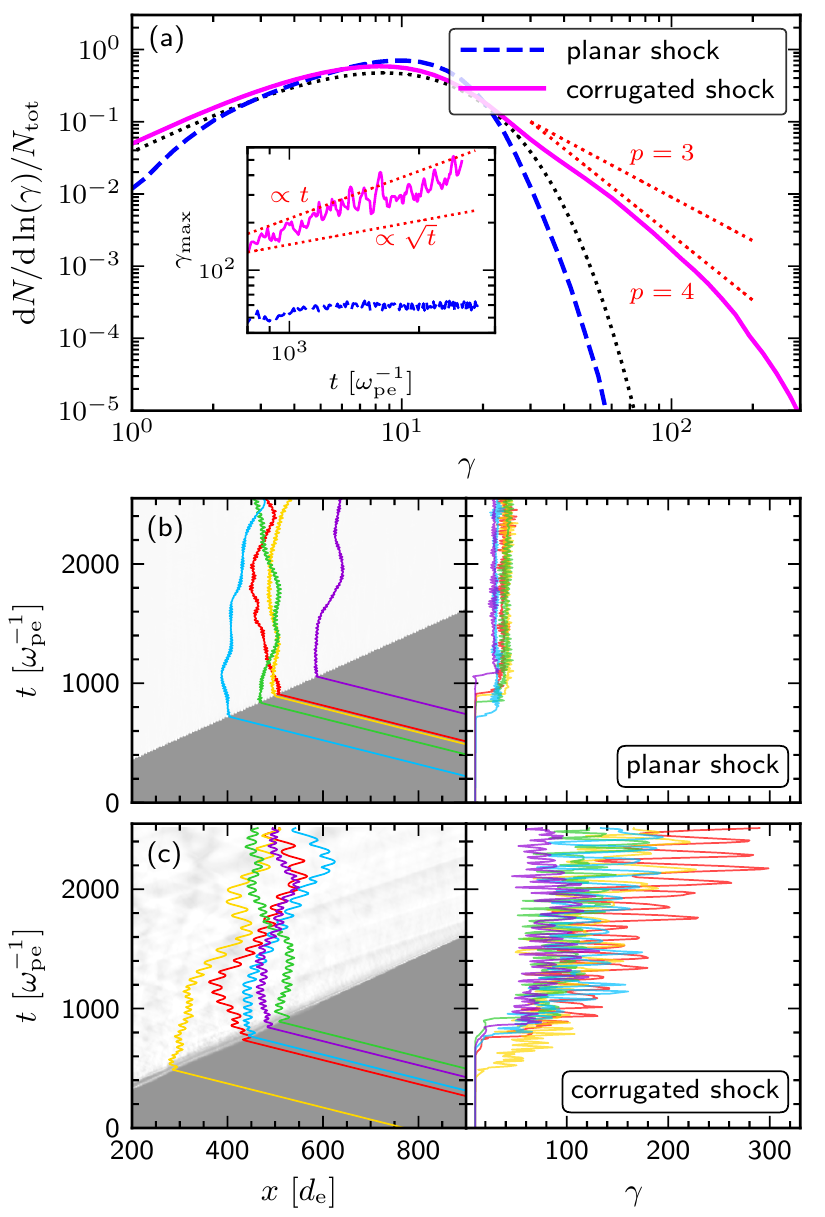}
    \caption{(a) Particle spectra for the fiducial setups with (solid line) and without (dashed line) corrugation in a region of the downstream ($400\le x/d_{\rm e}\le900$), averaged over 20 datasets at times $2285\le \omega_\mathrm{pe}t \le 2510$.
    For reference, the red dotted lines indicate the scaling $\mathrm{d}N/\mathrm{d}\gamma \propto \gamma^{-p}$ with index $p=3,\,4$, and 
    the black dotted line shows the 2D Maxwell-J\"{u}ttner distribution corresponding to the temperature predicted by the Rankine-Hugoniot relations for a homogeneous upstream flow ($k_\mathrm{B}T\approx 4.27 m_\mathrm{e}c^2$). 
    The inset shows for the two setups the maximum particle Lorentz factor in the selected downstream region as a function of time; the dotted lines follow the scalings $\propto \sqrt{t}$ and $\propto t$. 
    (b) $x$-position and Lorentz factor of five representative energetic particles for the homogeneous upstream setup. The shades of gray in the $x-t$ plane stand for the $y$-averaged density. (c) Same as (b) for the inhomogeneous upstream setup. }
    \label{fig:spec_etrack_vsplane}
\end{figure}

\section{Particle distribution and acceleration} \label{sec:prtcl}
The excited downstream shear motions can lead to particle energization behind the shock front.
To investigate this, we saved the history of positions and momenta of 8 million particles.
In panel (a) of Figure \ref{fig:spec_etrack_vsplane}, we show the particle distributions for the planar case (dashed blue line) and corrugated case with $\lambda_y=16\,d_{\rm e}$ (solid magenta line) in a slice of the downstream medium.
For reference, we also plot the thermal distribution corresponding to the temperature predicted by the Rankine-Hugoniot relations (dotted black line).
We see that in the homogeneous case, the particle distribution is quasi-thermal, as reported in previous studies \citep{Langdon1988,13Sironi,18Plotnikov}, while in the case of the perturbed inflow, the distribution develops a nonthermal tail. 
For reference, we indicated the scalings $\mathrm{d}N/\mathrm{d}\gamma \propto \gamma^{-4}$ and $\mathrm{d}N/\mathrm{d}\gamma \propto \gamma^{-3}$, close to what we could expect to emerge from stochastic acceleration in turbulence of moderate amplitude and magnetization \citep{18Comisso,19Comisso}.
The inset in panel (a) reports the maximal Lorentz factor of the tracked particles in the slice considered, $\gamma_\mathrm{max}$, as a function of time. 
For the case of a homogeneous preshock medium, $\gamma_\mathrm{max}$ quickly reaches saturation at a value of about $50-60$. 
In contrast, when the shock interacts with upstream perturbations, we observe the growth of $\gamma_\mathrm{max}$.
We indicated for reference the scalings $\gamma_\text{max}\propto \sqrt{t}$ and $\gamma_\mathrm{max} \propto t$; longer simulations are needed for more precise measurements but the trend seems to be compatible with the Bohm regime.
At the end of the simulation ($t\omega_\mathrm{pe}\approx 2500$), the highest-energy particles have $\gamma_\mathrm{max}\approx 500$, and the energization process appears to be still ongoing.

Panels (b) and (c) of Figure \ref{fig:spec_etrack_vsplane} depict the time evolution of the $x$-position and Lorentz factor of five representative particles (the 10th, 30th, 100th, 1000th, and 3000th most energetic of the tracked particles at $t\approx2400\, \omega_\mathrm{pe}^{-1}$, in the same slice of the downstream), for the unperturbed and perturbed cases, respectively. 
In the laminar case, upon crossing the shock, particles encounter a stronger magnetic field (caused by the shock compression) while the (motional) electric field becomes negligible (in the downstream frame); they start depicting Larmor gyrations and are advected with the magnetic field lines they are attached to without getting a chance to cross the shock again. Particles thus only display a small jump in energy corresponding to thermalization as they cross the shock and conserve an approximately constant energy once in the downstream. In the corrugated case, the initial kick in energy as the particles encounter the shock front is larger by about a factor of 2. The deformation of the shock and the inhomogeneity of the electromagnetic fields behind the shock allow some particles to cross back the shock during their first Larmor gyration. This is notably the case for particles penetrating the shock front through low-density regions where the magnetic field is smaller and the corresponding Larmor radius larger than in the unperturbed case. Particles then gain energy from the upstream motional electric field as they gyrate in the same direction. 
Particles thus typically experience a shock-drift acceleration cycle before being advected in the downstream where they go through an additional secular energization.
This energization appears to proceed through a stochastic acceleration channel \cite[e.g.,][]{Lemoine2019,Lemoine2021,19Rieger,CD20}: 
the perturbations in the bulk velocity of the downstream region induce fluctuating motional electric fields, which cause the particles to experience positive and negative energy changes with a positive net effect. 
Particle tracking indicates that this acceleration process is important when the particle's Larmor radius is comparable to the length scale of the downstream fluctuations. 
The nonideal electric field component, $E_\parallel \approx E_z$, is found to be negligible compared to the magnitude of $E_\perp$, indicating that nonideal magnetohydrodynamic effects (e.g., magnetic reconnection) are not active in the downstream (even in full 3D simulations). 

\begin{figure}[t]
    \centering
    \includegraphics[width=0.47\textwidth]{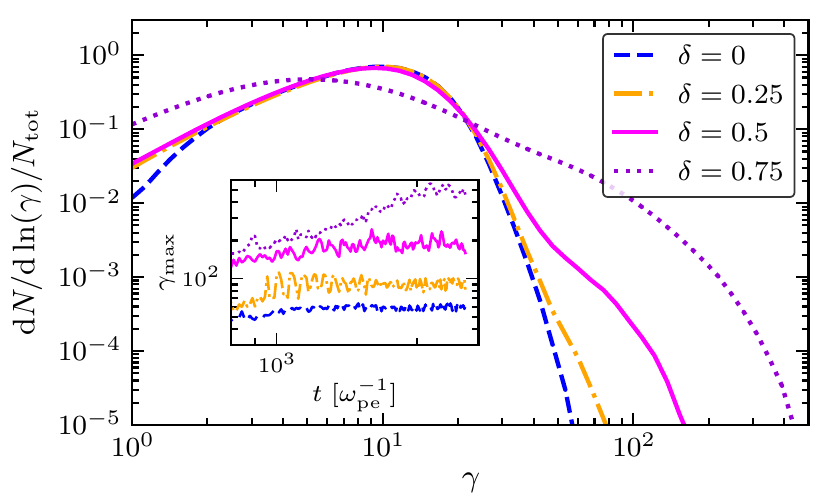}
    \caption{
    Particle spectra and maximum Lorentz factor in a region of the downstream ($400\le x/d_{\rm e}\le900$) for different incoming wave amplitudes $\delta$ and $\lambda_y=32\, d_\mathrm{e}$. 
    See panel (a) of Figure \ref{fig:spec_etrack_vsplane} and text for details.}
 \label{fig:ampli-dep}
\end{figure}

\begin{figure}[ht]
    \centering
    \includegraphics[width=0.47\textwidth]{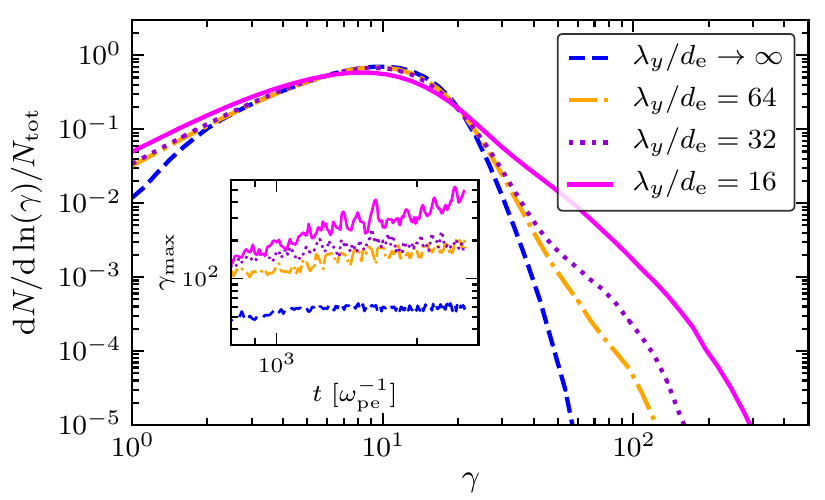}
    \caption{
    Particle spectra and maximum Lorentz factor in a region of the downstream ($400\le x/d_{\rm e}\le900$) for $\lambda_y/d_{\rm e}=16,\ 32, 64$ or $\to \infty$ (planar shock) and $\delta=0.5$. 
    See panel (a) of Figure \ref{fig:spec_etrack_vsplane} and text for details.}
 \label{fig:lamda-dep}
\end{figure}

To assess the generality of these results, we conducted simulations for different amplitudes (Figure \ref{fig:ampli-dep}) and wavelengths (Figure \ref{fig:lamda-dep}) of the upstream perturbations. 
We find that the effects of the upstream wave on the shock structure and the emergence of energetic particles become more important for increasing amplitudes. 
For $\delta= 0.25$, the particle energy distribution at $\omega_{\rm pe}t\approx 2400$ only slightly differs from that of the homogeneous case; even with this low amplitude setup, particles still experience a cycle of shock-drift acceleration accounting for the observed small excess of nonthermal particles and a saturation value about 30\% larger than in the planar case but the strength of the downstream fluctuations does not seem sufficient to allow significant stochastic acceleration within the timescale of the simulation.
For the case $\delta=0.75$ on the other hand, the spectrum displays a nonthermal power-law tail with index $p\approx2.5$ and the temporal evolution of the maximum energy is compatible with $\gamma_\mathrm{max}\propto t$ like one would expect in the Bohm regime.
Regarding the influence of $\lambda_y$, Figure \ref{fig:lamda-dep} suggests that the second stage of the acceleration process (taking place in the downstream) is more efficient for smaller incoming wavelengths. 
It may be that the larger dynamic range of the fluctuations excited downstream translates into a smaller injection energy for the acceleration process.

Finally, we note that the spectra and evolution of $\gamma_\mathrm{max}$ showed above were all computed in the region $400\le x/d_{\rm e}\le900$ of the downstream. We observed that the nonthermal tail becomes more important (in terms of both number fraction and energy cutoff) as the selected slab is moved further away from the shock, which is expected since particles have had more time to be accelerated. The slab selected to make Figures \ref{fig:spec_etrack_vsplane}--\ref{fig:lamda-dep} limits the risk of contamination by potential unphysical effects caused by the boundary conditions and/or early stages of the plasma reflection and shock formation. 
Larger and longer simulations are needed to study the shock distance dependence of particle energization.

\section{Discussion}
In this Letter, we investigated the properties of relativistic magnetized perpendicular shocks propagating into inhomogeneous environments.
By means of 2D and 3D PIC simulations, we have provided evidence that these corrugated shocks can lead to the generation of energetic particles---in contrast to previous studies of planar shocks.
Our study shows that corrugation induces motional electric fields in the downstream associated with a secular energization of particles.
This energization process is similar to the shear-driven acceleration mechanism observed in \cite{20Cerutti} and \cite{20Sironi} with the interesting difference that in our case, the physics of injection into the acceleration process is not mediated by magnetic reconnection.
This opens up the prospect of investigating from first-principles stochastic particle acceleration in a new regime outside of the two-stage acceleration paradigm (reconnection followed by stochastic acceleration) that has emerged from recent PIC simulations \citep{18Comisso,19Comisso,NattilaBeloborodov21}. 

This work has addressed the general problem of an upstream medium carrying density perturbations, with an unspecified origin, described as plane waves, whose polarization and amplitude were free parameters.
We focused here on results obtained with parameters associated with appreciable downstream perturbations ($\delta \rho/\rho\sim 1$, $k_x\ll k_y$) so that particle acceleration could be observed within the physical timescales currently achievable by our simulations.
In an astrophysical context, one could envision these small-scale inhomogeneities as the result of plasma instabilities such as the filamentation instability \citep[e.g.,][]{22Sobacchi, 23Sobacchi}, local density enhancements generated by the charge-starvation of Alfv\'en waves  \citep{Chen2022, NattilaBeloborodov2022}, or the formation of coherent structures (such as current sheets) in magnetic turbulence \citep[e.g.,][]{19Comisso}.

Efficient particle acceleration in the downstream requires an injection mechanism and strong motional electric fields, i.e., strong perturbations of the downstream magnetic and/or velocity fields, which in turn implies that the shock must be significantly deformed ($\delta X_\text{sh}/\lambda_y\sim 1$) by the upstream perturbations.
Our simulations also reveal that the particle acceleration efficiency has a complex (joint) dependence on the perturbation amplitude $\delta$ and wavelength $\lambda$.
Over the timescales that we probed, particle acceleration appears to be mostly resonant and therefore to be the most efficient when the wavelength of the upstream fluctuations is comparable to the Larmor radius of the particles in the downstream; for the parameters considered in this Letter, this corresponds to $\lambda_y \sim 10\, d_\mathrm{e}$. For perturbations on larger lengthscales, nonresonant particle acceleration \citep[e.g.,][]{22Bresci} probably plays a more important role, if particles still experience some injection mechanism.

We observe robust acceleration in the downstream down to amplitudes of $\delta \approx 0.5$. 
Here, $\delta = 0.5$ translates into decaying downstream perturbations with mildly relativistic bulk motions and $\sqrt{\langle \delta B^2 \rangle}/B_0 \approx 0.1$ which is, to our knowledge, the lowest amplitude probed by kinetic simulations of relativistic turbulence so far \citep{18Comisso}.
Probing the regime of small amplitudes, which are more relevant to critically balanced astrophysical turbulence ($\delta \ll 1$), and/or large wavelengths  ($\lambda/d_\mathrm{e} \gg 100$) would therefore require prohibitively long shock simulations. 
Constraints on the numerical cost also motivated the usage of 2D waves to reduce the dimensionality of the problem, as 3D simulations quickly become computationally infeasible.

We also note that in this Letter, we chose to parameterize the amplitude of the upstream perturbations in terms of $\delta=\delta \rho/\rho$ as it is well suited for the case of density waves in a moderately magnetized medium. A more general proxy of the degree of perturbation of the upstream medium that can be used when considering other types of waves or magnetizations is the relative variation of the energy-momentum flux induced by the perturbation \citep{18CD}. 
For highly magnetized media ($\sigma \gtrsim 1$), $\delta\sim1$ density waves only induce a small perturbation of the energy-momentum flux going through the shock resulting in inefficient downstream flow perturbation and particle acceleration. Magnetic perturbations are therefore more relevant for the regime of high magnetizations and we anticipate that magnetic reconnection plays an important role in the injection physics for that regime \citep{13Sironi,20Cerutti}.
A more comprehensive study investigating how the physics of injection and efficiency of particle acceleration depend on different physical parameters and geometries is thus certainly needed and deferred to future work.


Finally, an important feature of corrugated shocks that our simulations reveal is the formation of coherent, large-scale magnetized structures in the downstream region.
This is particularly important in light of the high X-ray polarization found in blazar Mrk~501 \citep{Liodakis2022}, which requires ordered magnetic field at the site of particle acceleration.
These results appeal to reconsider relativistic shocks as plausible sites for particle acceleration in magnetized astrophysical environments, such as relativistic jets.

\section*{Acknowledgments}
We thank the anonymous referee for their valuable comments.
The computations were enabled by resources provided by the Swedish National Infrastructure for Computing (SNIC) at the PDC Center for High Performance Computing, KTH Royal Institute of Technology.
Simons Foundation is also acknowledged for computational support. 
C.D. acknowledges support from NSF grant AST 1903335 and NASA grant NNX17AK55G.
J.N. is supported by a joint Columbia University/Flatiron Research Fellowship.
A.V. acknowledges the Academy of Finland grant 309308.
Our work also benefited from discussions during Team Meetings in the International Space Science Institute (Bern).
Nordita is supported in part by Nordforsk.

\bibliography{refs}{}
\bibliographystyle{aasjournal}

\end{document}